\documentclass[aps,prb,superscriptaddress,twocolumn,showpacs]{revtex4-1}
\usepackage{amsmath}	% required for `\align' (yatex added)
\usepackage{txfonts,graphicx,wasysym,bm,color,ulem}

\newcommand{\mysection}[1]{\section{#1}}

\renewcommand{\vec}[1]{{\bm #1}}

\begin{document}

%\preprint{}

% \title{Chiral condensate as a possible spin-unpolarized $\nu=0$
% quantum Hall state in graphene}
\title{Spin-resoloved chiral condensate as a spin-unpolarized $\nu=0$ quantum Hall state
in graphene}

\author{Yuji Hamamoto}
%\thanks{}
\altaffiliation{Present address: Department of Precision Science and Technology,
Osaka University, Suita 565-0871, Japan}
\affiliation{Institute of Physics, University of Tsukuba, Tsukuba 305-8571, Japan}

\author{Tohru Kawarabayashi}
\affiliation{Department of Physics, Toho University, Funabashi 274-8510, Japan}

\author{Hideo Aoki}
\affiliation{Department of Physics, University of Tokyo, Tokyo 113-0033, Japan}

\author{Yasuhiro Hatsugai}
\email[]{hatsugai.yasuhiro.ge@u.tsukuba.ac.jp (corresponding author)}
\affiliation{Institute of Physics, University of Tsukuba, Tsukuba 305-8571, Japan}
\affiliation{Tsukuba Research Center for Interdisciplinary Material Science,
University of Tsukuba, Tsukuba 305-8571, Japan}


\date{\today}

\begin{abstract}
Motivated by the recent experiments indicating a spin-unpolarized
$\nu=0$ quantum Hall state in graphene,
we theoretically investigate the ground state 
based on the many-body problem projected onto the $n=0$ Landau level.
For an effective model with the on-site Coulomb repulsion and antiferromagnetic
exchange couplings,
we show that the ground state is a doubly-degenerate {\it spin-resolved chiral condensate}
in which all the zero-energy states with up spin are condensed into one chirality,
while those with down spin to the other.
This can be exactly shown for an Ising-type exchange interaction.
%, and is adiabatically continued to the isotropic case.  
The charge gap due to the on-site repulsion in the ground state is shown to 
grow linearly with the magnetic field, 
in qualitative agreement with the experiments.

\end{abstract}


%\keywords{}

\maketitle

\mysection{Introduction}%
One of the most typical features of graphene
is the quantum Hall effect with quantized Hall plateaus
at filling factors $\nu=\pm2,\pm6,\pm10,\cdots$, a sequence that 
hallmarks Dirac electrons in magnetic fields.  Then we can pose a 
question: is there anything special occuring right at the Dirac point 
(at which the Landau level filling is $\nu=0$)?    
Soon after the observation of the quantum Hall sequence,
experiments have indeed 
discovered new conductivity plateaus at $\nu=0,\pm1,\pm4$
for strong enough 
magnetic fields.~\cite{PhysRevLett.96.136806,PhysRevLett.99.106802}
The new plateaus have naurally been drawing considerable theoretical 
attention.~\cite{PhysRevLett.96.256602,
PhysRevB.74.075422,
PhysRevB.74.161407,
PhysRevB.74.195429,
PhysRevLett.98.016803,
PhysRevB.75.165411,
2007SSCom.143..504A,
PhysRevLett.99.196802,
2008PhyE...40.1530H,
PhysRevB.80.235417,
PhysRevLett.103.216801,
PhysRevB.81.075427,
PhysRevB.81.115405,
PhysRevB.81.205429,
PhysRevB.85.155439,
PhysRevB.86.205424}
A particular interest is this might be a manifestation 
of many-body effects in graphene, which is an unusually clean system.  
Specifically, special attention has been paid to the $\nu=0$ situation, 
where experiments have observed unusual behaviors distinct from other fillings.
Namely, the $\nu=0$ state exhibits an unexpected insulating behavior
with exponentially diverging longitudinal resistivity, 
which suggests that the system undergoes a Mott transition
at half filling.~\cite{PhysRevLett.100.206801}
Moreover, recent experiments
 on high quality samples on hBN substrates
have revealed a spin-unpolarized aspect of the $\nu=0$ state, 
along with a suggestive energy gap growing linearly
with the perpendicular magnetic field $B$.~\cite{PhysRevLett.108.106804,2012NatPh...8..550Y}
The latter finding should provide an important clue to the theoretical understanding
of the $\nu=0$ state,
since the linear $B$-dependence is incompatible with a naive estimation
based on the Dirac field model in continuum space (as opposed to the 
honeycomb lattice model), 
where a many-body gap due to the Coulomb interaction should scale as $e^2/l_B\propto\sqrt{B}$ with $l_B=\sqrt{\hbar/eB}$ being the magnetic length.
While it has been proposed that the lattice effect
leads to a linear dependence of the gap.~\cite{2007SSCom.143..504A,
PhysRevB.74.075422,PhysRevLett.99.196802} 
the spin-unpolarized nature of the $\nu=0$ state has yet to be fully understood.

This has motivated us here to theoretically investigate the spin-unpolarized $\nu=0$ state
with a special emphasis on the chiral symmetry. The symmetry is indeed 
a fundamental aspect of the graphene honeycomb lattice, 
and plays a crucial role in the peculiar electronic properties of graphene 
already in the one-body problem.   Namely, 
the doubled Dirac cones are guaranteed by the chiral symmetry, 
which can be called a two-dimensional analog of the Nielsen-Ninomiya's theorem
in the (3+1)-dimensional gauge theory. 
In a perpendicular magnetic field, the chiral symmetry 
affects most remarkably the $n=0$ Landau level (LL), 
where the $\delta$-function-like density of states is topologically protected
even in disordered systems as long as the disorder respects the chiral
symmetry.~\cite{PhysRevLett.103.156804}
The chiral symmetry should also exert important effects for many-body problems in the $n=0$ LL.  This is because we can characterize many-body states 
by the chiralities of filled zero modes.  
For a spin-split $n=0$ LL, 
the ground state is exactly shown to be a chiral condensate doublet
with a finite energy gap.~\cite{PhysRevB.86.205424,1367-2630-15-3-035023}
While the total Chern number for the chiral condensate is zero, because 
the contribution 
from the Dirac sea (negative-energy states) cancels the zero-mode Chern 
number, 
its topological nature is shown to appear as edge states with a characteristic bond order, 
which can be considered as an example of
the bulk-edge correspondence in topological systems.~\cite{PhysRevLett.71.3697}
In this Letter, we shed light on 
the spin-unpolarized 
nature of the $\nu=0$ state,
by extending the picture of the chiral condensate to accommodate 
the spin degree of freedom.  Based on a lattice model 
with on-site repulsive interaction and also a 
nearest-neighbor exchange coupling, 
the many-body ground state is shown to be a doubly-degenerate
{\it spin-resolved chiral condensate},
in which all the zero-energy states
with up spin are condensed into one chirality, while those with down spin to the other.
We have shown this exactly for an Ising-type 
exchange interaction, which is adiabatically continued to the isotropic case.  
The charge gap due to the on-site repulsion in the ground state turns 
out to grow linearly with the magnetic field, 
in qualitative agreement with the experiments.~\cite{2012NatPh...8..550Y}

\mysection{Projection onto the $n=0$ Landau level}%
To describe the many-body problem in the $n=0$ LL,
we consider a projected Hamiltonian, $\tilde{H}=P(H_t+H_U+H_J)P^{-1}$, 
with $P$ denoting the projection onto the $n=0$ LL.
The kinetic part is given by a tight-binding Hamiltonian,
\begin{gather}
H_t=-t\sum_{\langle ij\rangle}\sum_{s=\uparrow\downarrow}
e^{i\theta_{ij}}c^\dagger_{is}c_{js}+{\rm H.c.},\label{eq:kinetic}
\end{gather}
where $t>0$ is the hopping between nearest-neighbor sites $\langle ij\rangle$, 
and $c^\dagger_{is}$ creates an electron with spin $s$ at $i$. 
The perpendicular magnetic field is introduced with the Peierls phase $\theta_{ij}$,
which is chosen so that the magnetic flux piercing a unit hexagon equals
$\phi=\frac{1}{2\pi}\sum_{\hexagon}\theta_{ij}$ in units of the flux quantum $\phi_0=h/e$.
For a torus geometry with $N$ unit cells,
the flux in the string gauge~\cite{PhysRevLett.83.2246} 
(which enables us to treat smaller fields) 
reads $\phi=M/N$ with an integer $M$.

We then turn on electron-electron interactions,
whose leading contribution is the on-site interaction,
% \red{fully taking into account the screening effect due to a dielectric substrate.
% Then the leading contribution comes from the on-site interaction},
\begin{gather}
 H_U=U\sum_ic^\dagger_{i\uparrow}c_{i\uparrow}c^\dagger_{i\downarrow}c_{i\downarrow},
\label{eq:on-site}
\end{gather}
with a repulsion $U>0$.
% In a magnetic field we have Landau's quantization,
% so that the kinetic energy is quenched in the $n=0$ LL.
% Since we have thus an infinitely strongly correlated system,
% we cannot proceed as e.g. in the ordinary Hubbard model with an expansion in $t/U$.
% However, entirely apart from such an expansion,
% there should be an exchange interaction,
% \begin{gather}
% H_J=J\sum_{\langle ij\rangle}\left[\alpha(S^x_iS^x_j+S^y_iS^y_j)+S^z_iS^z_j
% -\frac{1}{4}n_in_j\right],\label{eq:exchange}
% \end{gather}
% arising from a Coulomb matrix element as a next leading interaction after $U$.
% So we can put it in the interaction,
% while its magnitude can be calculated from first principles
% in terms of graphene Landau wave functions if so desired.
Matrix elements of the (direct and exchange) 
Coulomb interaction, on the other hand,
strongly depend on the LL index, where 
the short-range part is dominant in the $n=0$ LL.
Moreover, the long-range part of the interaction should be screened
on an ultraflat hBN substrate.  
Thus we include only the dominant nearest-neighbor interaction
in the form of an exchange interaction,
\begin{gather}
H_J=J\sum_{\langle ij\rangle}\left[\alpha(S^x_iS^x_j+S^y_iS^y_j)+S^z_iS^z_j
-\frac{1}{4}n_in_j\right],\label{eq:exchange}
\end{gather}
whose physical meaning is discussed below.
As we shall see, this acts to lift the degeneracy in the multiplet,
resulting in a spin-unpolarized ground state.
In Eq.~(\ref{eq:exchange}),
the factor $\alpha$ tunes the anisotropy in the exchange interaction,
varying between the Ising ($\alpha=0$) and the spherical ($\alpha=1$) limits.
We ignore the Zeeman effect, since it is much smaller than the other energy scales.

To derive the effective Hamiltonian in the $n=0$ LL,
we first diagonalize the kinetic term, Eq.~(\ref{eq:kinetic}).
Due to the chiral symmetry, $\{H_t,\Gamma\}=0$ 
with $\Gamma$ being the chiral operator,
a one-body state $\psi_\varepsilon$ at energy $\varepsilon$
is related to its chiral partner as $\psi_{-\varepsilon}=\Gamma\psi_{\varepsilon}$.
Thus a special situation arises in the $n=0$ LL,
where particle- and hole-states are degenerate.
As a result, there appears $2M$ zero modes in the string gauge.
By reconfiguring these zero modes, one obtains a chiral basis,
\begin{gather}
 \psi=(\psi_{1+},\cdots,\psi_{M_++},\psi_{1-},\cdots,\psi_{M_--}),\label{eq:basis}
\end{gather}
where $\{\psi_{k\pm}\}$ with $k=1,\cdots,M_\pm$ are eigenstates of the chiral operator 
satisfying $\Gamma\psi_{k\pm}=\pm\psi_{k\pm}$.
$M_\pm$ is the degeneracy of the zero modes with chirality $\pm$, 
hence $M_++M_-=2M$.  
While the kinetic energy is quenched in the $n=0$ LL,
the information on the kinetic part is encoded
in the properties of the chiral zero modes. 
A simplest example is the fact that 
chirality designates the sublattice on which a zero mode resides, 
i.e., $\psi_{k+(-)}$ has nozero amplitudes only on sublattice
$\bullet(\circ)$.~\footnote{In this sense,
chirality is analogous to valley pseudospin for the low-energy effective 
model, although in a magnetic field the Dirac cones coalesce into the LLs.}
In fact, this is a key to an exact treatment of the ground state
as we shall see.  

In terms of the chiral basis~(\ref{eq:basis}), 
the projection onto the $n=0$ LL is defined by a mapping
$c^\dagger_{is}\mapsto\tilde{c}^\dagger_{is}\equiv(c^\dagger_s\psi\psi^\dagger)_{i}$,
with a row vector $c_s^\dagger=(c^\dagger_{1s},\cdots,c^\dagger_{2Ns})$
and a projection matrix $\psi\psi^\dagger$.
Note that $\tilde{c}^\dagger_{is}$ no longer obeys the canonical anticommutation relations,
since the chiral basis Eq.~(\ref{eq:basis}) is not complete. 
Alternatively, we can introduce creation operators of the zero modes, 
$d^\dagger_{ks\pm}\equiv c^\dagger_s\psi_{k\pm}$, which satisfy the anticommutation relations
\begin{gather}
\{d_{ks\chi},d^\dagger_{ls'\chi'}\}
=\delta_{kl}\delta_{ss'}\delta_{\chi\chi'},\\
\{d_{ks\chi},d_{ls'\chi'}\}
=\{d^\dagger_{ks\chi},d^\dagger_{ls'\chi'}\}=0. 
\end{gather}
With these fermions we can rewrite the projected Hamiltonian as
$\tilde{H}=\tilde{H}_U+\tilde{H}_J$ with
\begin{gather}
\tilde{H}_U=\sum_{klmn}\sum_{\chi=\pm}\mathcal{U}_{klmn}^\chi
d^\dagger_{k\uparrow\chi}d^\dagger_{l\downarrow\chi}d_{m\downarrow\chi}d_{n\uparrow\chi},\\
\tilde{H}_J=\sum_{klmn}\sum_{s=\uparrow\downarrow}
\frac{\mathcal{J}_{klmn}}{2}
d^\dagger_{ks+}d^\dagger_{l\bar{s}-}
(\alpha d_{ms-}d_{n\bar{s}+}-d_{m\bar{s}-}d_{ns+}),\label{eq:exchange-projected}
\end{gather}
where $\bar{s}=\uparrow(\downarrow)$ for $s=\downarrow(\uparrow)$
and the pseudopotentials are defined as
\begin{gather}
\mathcal{U}_{klmn}^\pm
=U\sum_{i}(\psi_{k\pm})^\ast_i
(\psi_{l\pm})^\ast_i(\psi_{m\pm})_i(\psi_{n\pm})_i,\\
\mathcal{J}_{klmn}
=J\sum_{\langle i\in\bullet,j\in\circ\rangle}
(\psi_{k+})^\ast_i(\psi_{l-})^\ast_j(\psi_{m-})_j(\psi_{n+})_i.\label{eq:pseudo-exchange}
\end{gather}
From this form we can identify the meaning of the $J$ term: 
In a magnetic field we have Landau's quantization,
so that the kinetic energy is quenched in the $n=0$ LL.
We then end up with an infinitely strongly correlated system, 
so that we cannot proceed as e.g. in the ordinary Hubbard model with an expansion in $t/U$
arising from a Coulomb matrix element as a next leading interaction after $U$.
However, an exchange interaction between Landau basis functions 
should exist, whose magnitude can be calculated from first principles
in terms of graphene Landau wave functions if so desired.  We can 
thus interpret $J$ introduced in Eq.~(\ref{eq:exchange}) as representing the 
exchange interaction in Eq.~(\ref{eq:pseudo-exchange}).

When a many-body state is constructed by occupying the chiral zero modes, 
the {\it total chirality} is conserved, since $\tilde{H}$ commutes with the operator,
\begin{gather}
 \mathcal{G}=\sum_{s=\uparrow\downarrow}
\left(\sum_{k=1}^{M_+}d^\dagger_{ks+}d_{ks+}
-\sum_{k=1}^{M_-}d^\dagger_{ks-}d_{ks-}\right).
\end{gather}
This enables us to diagonalize $\tilde{H}$ separately 
in a subspace for each sector in the total chirality.

\mysection{Spin-resolved chiral condensate}%
To discuss the many-body problem,
the exchange interaction with an Ising anirotropy is a useful starting point 
for elucidating the true ground state.
At half filling, the projected Hamiltonian for $\alpha=0$
is rewritten, up to a constant, as
\begin{align}
 \tilde{H}&=\frac{U}{2}\sum_i\tilde{c}^\dagger_{i\uparrow}\tilde{c}^\dagger_{i\downarrow}
\tilde{c}_{i\downarrow}\tilde{c}_{i\uparrow}
+\frac{J}{4}\sum_{\langle ij\rangle}\sum_s
\tilde{c}^\dagger_{is}\tilde{c}_{j\bar{s}}\tilde{c}^\dagger_{j\bar{s}}\tilde{c}_{is}
+{\rm C.c.},\label{eq:ham-ising}
\end{align}
which is invariant for the charge conjugation (C.c.), 
$\tilde{c}_{is}\leftrightarrow\tilde{c}_{is}^\dagger$.
Since the Hamiltonian~(\ref{eq:ham-ising}) is semi-positive definite
$\langle \tilde{H}\rangle\ge 0$,
a state destructed by $\tilde{H}$
is the ground state for the system.
Such a ground state can be constructed as a doubly-degenerate chiral condensate,
\begin{gather}
 |G_{s\bar{s}}\rangle=\prod_{k=1}^{M_+} d^\dagger_{ks+}
\prod_{l=1}^{M_-} d^\dagger_{l\bar{s}-}|D_<\rangle\qquad
(s=\uparrow,\downarrow),\label{eq:chiral-condensate}
\end{gather}
where $|D_<\rangle$ denotes the Dirac sea of the negative energy states.
In Eq.~(\ref{eq:chiral-condensate}),
the zero modes with up-spin form a chiral condensate with chirality $+(-)$,
while those with down-spin a chiral condensate with chirality $-(+)$.
From the correspondence between the chirality and sublattices,
we can readily check that $|G_{\uparrow\downarrow}\rangle$ and
$|G_{\downarrow\uparrow}\rangle$ are indeed destructed by
$\tilde{c}_{i\downarrow}\tilde{c}_{i\uparrow},
\tilde{c}^\dagger_{j\bar{s}}\tilde{c}_{is}$ and their charge conjugates
in Eq.~(\ref{eq:ham-ising}).
If we restrict ourselves to the case of $M_+=M_-$,
which holds when the two sublattices contain the same number of sites,
the ground state falls upon the sector of total chirality
$\chi_{\rm tot}\equiv\langle\mathcal{G}\rangle=0$, 
in sharp contrast to the spinless case,~\cite{PhysRevB.86.205424,1367-2630-15-3-035023}
where the ground state is a chiral condensate with fully polarized chirality.
Although $|G_{s\bar{s}}\rangle$ forms a lattice-scale staggered spin order
in the $n=0$ LL,
the ground state is not a simple N\'eel state,
since the two chiral condensates form a doublet
$\Psi=(|G_{\uparrow\downarrow}\rangle,|G_{\downarrow\uparrow}\rangle)$
even for a finite system, 
and can be mixed through a unitary transformation $\Psi=\Psi_\omega\omega$
with $\omega\in U(2)$.
Note that since the chiral condensate has no double occupancy on a site,
it can be considered as the ground state for the $t$-$J$ model,
which coincides with the strong $U$ limit of the present model.

\begin{figure}[t]
\includegraphics[width=.86\linewidth]{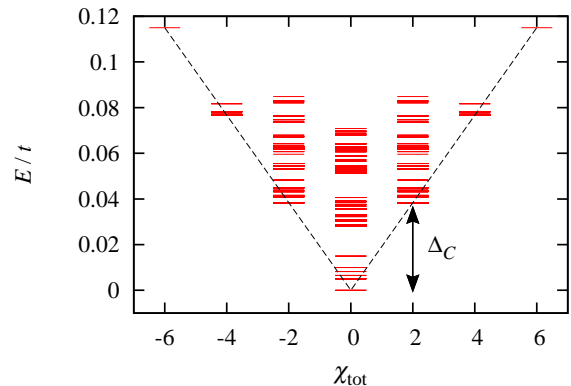}
\caption{\label{fig:spec}
(Color online) Energy spectrum 
classified according to the total chirality $\chi_{\rm tot}$ 
 in the Ising limit $\alpha=0$, for 
$M=3, \phi=1/300, U/t=10$ and $J/t=1$.
The spectrum is symmetric about $\chi_{\rm tot}=0$, 
and the bottoms of different sectors exhibit a linear increase 
with $|\chi_{\rm tot}|$ as indicated by dashed lines.
}
\end{figure}

The excited states above the ground state can be obtained
by numerically diagonalizing the projected Hamiltonian $\tilde{H}$.
In Fig.~\ref{fig:spec}, we show the energy spectrum in the Ising limit $\alpha=0$
for $\phi=1/300$, $M=3$, $U/t=10$ and $J/t=1$.
Here we have classified the spectrum according to the total chirality $\chi_{\rm tot}$, 
which takes even numbers as $\chi_{\rm tot}=0,\pm2,\pm4,\cdots,\pm 2M$.
Let us first focus on the sector of $\chi_{\rm tot}=0$,
where the chiral condensate~(\ref{eq:chiral-condensate})
is indeed obtained as the doubly-degenerate ground state
as expected from the above discussion.
For $J\ll U$, the low-energy excitations in the central sector
are created by spin flipping,
so that the Ising anisotropy opens a finite gap above the ground state.
This makes the Chern number of the chiral condensate doublet well-defined
and thereby allows us to calculate the Hall conductivity 
with the Niu-Thouless-Wu formula,~\cite{PhysRevB.31.3372}
\begin{gather}
 \sigma_{xy}=\frac{e^2}{h}\frac{1}{N_D}C,\qquad
C=\frac{1}{2\pi i}\int{\rm Tr}dA,\qquad
\end{gather}
where $N_D=2$ is the ground state degeneracy, and
$A=\Psi^\dagger d\Psi$ is the non-Abelian Berry connection
for multiplets.~\cite{2004JPSJ...73.2604H}
Since the Hall conductivity does not distinguish the spin degree of freedom,
the Chern number of the chiral condensate trivially
doubles the result in the spinless case.~\cite{PhysRevB.86.205424,1367-2630-15-3-035023}
Thus, from the sum rule for the Chern number,
the Hall conductivity is analytically calculated as $\sigma_{xy}=0$,
which corresponds to the Hall plateau at zero around the half filling
$\nu=0$.~\cite{PhysRevLett.96.136806}

\mysection{Charge gap and the spherical limit}%
If we now turn to the other sectors of $\chi_{\rm tot}$ in the energy 
spectrum, 
we immediately notice that the entire picture of the spectrum
has a reflectional symmetry with respect to $\chi_{\rm tot}=0$,
which reflects the invariance of $\tilde{H}$ against global chirality flipping.
More importantly, the bottoms of different sectors delineate a linear 
increase with $|\chi_{\rm tot}|$, as indicated with the dashed lines in Fig~\ref{fig:spec}, which is a key result in the present work.  
This can be understood by considering the on-site repulsion between the zero modes.
Since all the zero modes are singly occupied in the ground state~(\ref{eq:chiral-condensate}),
single flips in the chirality inevitably involve a double occupancy of zero modes,
which opens a gap $\Delta_C$ in the neighboring sector.
The behavior of the charge gap becomes clearer
by taking a closer look at the lowest-energy states in the sector of $\chi_{\rm tot}=\pm2$.
The degeneracy of them is numerically determined to be $4M^2$,
which suggests that they can be written as
\begin{gather}
|E^{kl}_{ss'\chi}\rangle=d^\dagger_{ks',-\chi}d_{ls'\chi}|G_{s\bar{s}}\rangle,
\qquad
(s',\chi)=(s,+),(\bar{s},-)\label{eq:excited}
\end{gather}
for various zero-mode indices $k$ and $l$.
Note that this is reminiscent of the projected single-mode
approximation.~\cite{PhysRevLett.54.581,PhysRevB.33.2481,PhysRevLett.73.3568}
Using Eq.~(\ref{eq:excited}), we can {\it analytically} obtain
the charge gap as
\begin{gather}
\Delta'_C\equiv\langle E_{ss'\chi}^{kl}|\tilde{H}|E_{ss'\chi}^{kl}\rangle
=\left(U+\frac{3}{2}J\right)\phi\label{eq:charge-gap}
\end{gather}
%\remove{where the magnetic flux $\phi$ stems from the fact that}
in the Landau gauge (see Appendix), where the chiral condensate has
a uniform local density of states,
$\langle G_{s\bar{s}}|\sum_{s'}\tilde{c}^\dagger_{is'}
\tilde{c}_{is'}|G_{s\bar{s}}\rangle=\phi$.
Within numerical error,
Eq.~(\ref{eq:charge-gap}) reproduces the numerical result for $\Delta_C$,
which is obtained from the difference between the ground energies
in the sectors of $\chi_{\rm tot}=0$ and $\chi_{\rm tot}=\pm2$.
Note that,
while a $\phi$-linear gap is obtained even for $J=0$ from Eq.~(\ref{eq:charge-gap}),
finite $J$ has been crucial for the exact treatment
of the spin-unpolarized ground state~(\ref{eq:chiral-condensate})
and the charge gap $\Delta'_C$.

\begin{figure}[t]
 \includegraphics[width=0.891\linewidth]{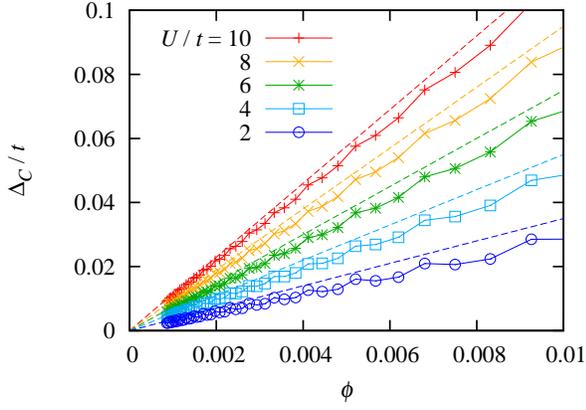}
\caption{\label{fig:gap-spherical}
(Color online) Charge gap $\Delta_C$ in the spherical limit $\alpha=1$ 
against the magnetic flux $\phi$ for 
$U/t$ varied from 2 to 10 with $M=3$ and $J/t=1$.  
For comparison, the analytic result, $\Delta'_C \propto \phi$,
in the Ising limit~(\ref{eq:charge-gap}) is also shown with dashed lines.}
\end{figure}

The charge gap is important in analyzing
the experimental results for the $\nu=0$ state.
Since at half filling an electric current has to be accompanied by double occupancies of lattice sites, 
the transport measurement should reflect the charge gap above the ground state.
More explicitly, the current operator defined in the projected subspace,
\begin{align}
 I_{ij}
&=i\sum_{kl}(\psi_{k+})_i^\ast(\psi_{l-})_j\sum_sd^\dagger_{ks+}d_{ls-}+{\rm H.c.},
\label{eq:current}
\end{align}
has nonzero matrix elements only between neighboring sectors of $\chi_{\rm tot}$,
while no electric current is carried by the low-energy excitations within one sector. 
Experimentally, the energy gap observed at $\nu=0$ displays
a linear dependence on the magnetic field $B$,~\cite{2012NatPh...8..550Y}
rather than a dependence, $e^2/l_B\propto\sqrt{B}$, for a 
long-range Coulomb interaction.  
Thus the charge gap $\propto B$ for the chiral condensate~(\ref{eq:charge-gap}) 
agrees qualitatively with the experiments.  

\begin{figure}[t]
 \includegraphics[width=0.891\linewidth]{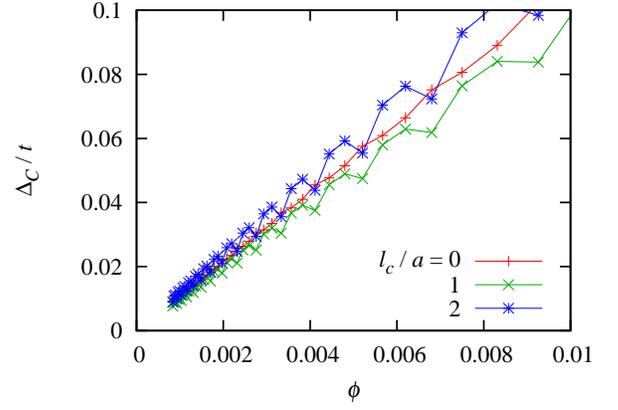}
\caption{\label{fig:gap-long-range}
(Color online) Influence of finite-range interactions on $\Delta_C$ is shown for
$U/t=10, V/t=1$ and various values of the cutoff distance, $l_c$,
in the off-site interaction.
The other parameters are the same as in Fig.~\ref{fig:gap-spherical}.}
\end{figure}

Next we move on to
%\remove{an obvious}
a natural question of what happens when the Ising 
anisotropy is made spherical.  In this case 
spin flipping occurs in the exchange Hamiltonian~(\ref{eq:exchange-projected}). 
In Fig.~\ref{fig:gap-spherical}, we plot the result for 
$\Delta_C$ against  $\phi$ in the spherical limit $\alpha=1$, 
where $U/t$ is varied from 2 to 10
and the other parameters are the same as in Fig.~\ref{fig:spec}.
We can see that the gap still grows approximately linearly
with $\phi$.~\footnote{A slight, 
triply-periodic oscillation in the data against $\phi$ 
is an effect of finite-range interactions on the honeycomb 
lattice, and becomes negligible for $U\gg J$.}
This suggests that the linear $B$ dependence essentially derives from the on-site repulsion, and does not depend on the detail of the exchange interaction.  
Note that the charge gap in Fig.~\ref{fig:gap-spherical} is slightly smaller 
than the Ising result [Eq.~(\ref{eq:charge-gap}); the dashed lines], 
since the spin flipping in Eq.~(\ref{eq:exchange-projected}) decreases the exchange energy.  
Assuming $U=10$ eV and $J=5$ eV in Eq.~(\ref{eq:charge-gap}), 
we can estimate the charge gap to be $\Delta_C$~[K]~$\sim 2.6B$~[T].  
The linear $B$ dependence agree with the experimental results,~\cite{2012NatPh...8..550Y}
although the size of the theoretical gap 
is smaller by a factor of 5.  
However, $B$-linear gap itself 
persists, as displayed in Fig.~\ref{fig:gap-long-range}, 
even when the on-site interaction [first term on the right-hand 
side of Eq.~(\ref{eq:ham-ising})] is made finite-ranged %\remove{as}
by adding
% $\tilde{H}_V = (1/2)\sum_{i\ne j}\sum_{ss'}V_{ij}\tilde{c}^\dagger_{is}\tilde{c}_{is}
% \tilde{c}^\dagger_{js'}\tilde{c}_{js'}$ with an off-site interaction 
% $V_{ij}=V/|\vec{i}-\vec{j}|$ for $|\vec{i}-\vec{j}|\le l_c/a$ and $V_{ij}=0$ otherwise,
\begin{gather}
 \tilde{H}_V = \frac{1}{2}\sum_{i\ne j}\sum_{ss'}V_{ij}\tilde{c}^\dagger_{is}\tilde{c}_{is}
\tilde{c}^\dagger_{js'}\tilde{c}_{js'}
\end{gather}
with an off-site interaction
\begin{gather}
 V_{ij}=\left\{
\begin{array}{cc}
 \frac{V}{|\vec{i}-\vec{j}|}&|\vec{i}-\vec{j}|\le \frac{l_c}{a}\\
0&\mbox{otherwise}
\end{array}
\right.,
\end{gather}
where $V>0$ and
$l_c/a$ is a cutoff in units of the inter-atomic distance $a\simeq 0.142$\,nm.
Thus the $B$ dependence is not restricted to the on-site 
interaction as long as $l_c<l_B$ and $V$ is sufficiently smaller than $U$.
Inclusion of long-range interactions beyond $l_B$ will be 
an intriguing extension
of the present problem, where it is expected that
the behavior of the gap would cross over to $\Delta_C\propto\sqrt{B}$
as observed in recent experiments in suspended 
(hence less screened) graphene.~\cite{PhysRevB.88.115407}

\begin{figure}[t]
 \includegraphics[width=\linewidth]{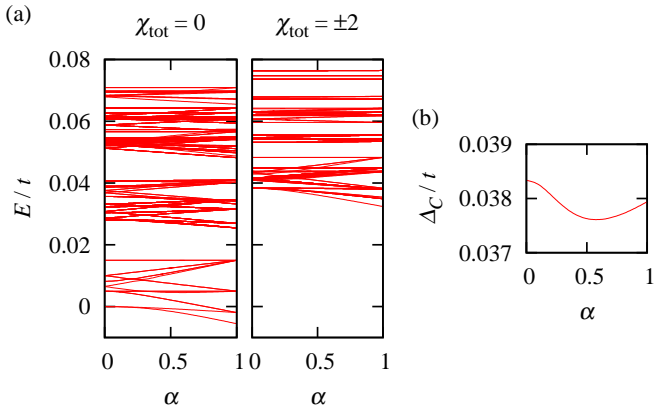}
\caption{\label{fig:adiabatic}
(Color online) $\alpha$-dependence of the energy spectrum
for the sectors of $\chi_{\rm tot}=0$ and $\chi_{\rm tot}=\pm2$ (a) and the charge gap (b).
The parameters are the same as in Fig.~\ref{fig:spec}.}
\end{figure}

Finally, we discuss how the chiral condensate~(\ref{eq:chiral-condensate}) 
evolves in the spherical limit $\alpha=1$. 
Exact diagonalization for $\alpha=1$ shows that the  ground state 
is spin-singlet, i.e., the ground state is spin-unpolarized in both 
the Ising and spherical limits in our model.  
This suggests that they are adiabatically connected when 
the value of $\alpha$ is varied.  
We have calcualted the adiabatic flow of the energy spectrum 
in Fig.~\ref{fig:adiabatic}, which shows that they are indeed connected.  
Namely, while the Ising gap above the chiral condensate closes at $\alpha=1$
for large systems,
the charge gap remains open irrespective of 
the anisotropy in the exchange coupling as shown in Fig.~\ref{fig:adiabatic}(b).
Thus, under the selection rule of Eq.~(\ref{eq:current}) 
which projects out the low-energy spin excitations,
the charge gap is adiabatically connected between the two limits.
The robustness of the charge gap suggests that the chiral condensate
captures the essence of the true ground state.

\mysection{Summary}%
We have theoretically investigated the spin-unpolarized aspect
of the $\nu=0$ quantum Hall state in graphene 
based on the many-body problem in the $n=0$ Landau level 
taking into account on-site repulsive interaction
and nearest-neighbor exchange interaction.
In the Ising limit of the exchange coupling,
the ground state is exactly shown to be a spin-resolved chiral condensate,
and the charge gap above the ground state grows linearly with the magnetic field.
The spin-unpolarized nature and the linear $B$ dependence of the charge gap
are retained when the exchange interaction is made isotropic, 
and the result qualitatively agrees with the recent
experiments.~\cite{PhysRevLett.108.106804,2012NatPh...8..550Y}

% \red{\mysection{Note added}%
% After submitting the manuscript,
% we became aware of Ref.~\cite{abanin-unpublished},
% where $\sqrt{B}$ dependence of the $\nu=0$ gap in suspended graphene
% is discussed by taking into account long-range Coulomb interaction.
% In this Letter, on the other hand,
% we have assumed that the long-range part is fully screened by a dielectric substrate.
% }

\acknowledgements

The work is supported in part by Grants-in-Aid for Scientific Research No. 23340112 from JSPS.
Ya.H. is also supported by No. 2561010, No. 25610101 and No. 23540460.
The computation in this work has been done with the facilities of the Supercomputer Center,
Institute for Solid State Physics, University of Tokyo. 

\appendix
\section*{Charge gap above the chiral condensate}
In this appendix, we analytically calculate the eigenenergy of
the excited state~(\ref{eq:excited}) to show that the charge gap
above the chiral condensate~(\ref{eq:chiral-condensate})
scales linearly with $\phi$.
To this end we first note that the chiral condensate has a uniform
local density of states (LDOS) around zero energy as
\begin{gather}
 \langle G_{s\bar{s}}|\sum_{s'}\tilde{c}^\dagger_{is'}\tilde{c}_{is'}|G_{s\bar{s}}\rangle
=(\psi\psi^\dagger)_{ii}\equiv n_0,
\end{gather}
with the projection matrix $\psi\psi^\dagger$.
This can be exactly shown in the Landau gauge,
where the system retains the translational and sublattice symmetries
in uniform magnetic fields.
The uniform value $n_0$ can be readily obtained as follows:
For a half-filled system composed of $N$ unit cells,
the electron density on a site equals $1/2N$ per state.
For a magnetic flux $\phi=M/N$ ($M$: integer)
the $n=0$ LL is $2M$-fold degenerate for each spin.
Thus the LDOS is equal to the flux as
\begin{gather}
 n_0=\frac{1}{2N}\cdot 2M=\phi.
\end{gather}
It should be noted that the string gauge~\cite{PhysRevLett.83.2246} 
enables us to investigate smaller magnetic fields 
than in the Landau gauge, but 
the translational symmetry is broken.  
While this implies that the LDOS is slightly dependent on 
the position $i$, 
the deviation is negligibly small 
in large systems, or equivalently, in small magnetic fields
treated in the numerical calculation in this paper.

Thus we calculate the eigenenergy of the excited state 
for the uniform LDOS $n_0=\phi$.
When the on-site and exchange Hamiltonians are operated on the excited state,
most terms vanishes due to the relations
$\tilde{c}_{i\downarrow}\tilde{c}_{i\uparrow}|G_{s\bar{s}}\rangle=0$, etc.,
and also to the fact that the chiral zero mode has nonzero amplitudes
only on one sublattice. 
Hence we have
\begin{align}
 \tilde{H}_U|E_{ss'\chi}^{kl}\rangle
&=U\sum_i
(\psi\psi^\dagger)_{ii}
(\psi_{k,-\chi})_i\tilde{c}^\dagger_{is'}d_{ls'\chi}
|G_{s\bar{s}}\rangle\\
&=U\phi|E_{ss'\chi}^{kl}\rangle,\label{eq:gap-on-site}\\
\tilde{H}_{J}|E_{ss'\chi}^{kl}\rangle
&=\frac{3}{4}J\sum_{i\in\bullet}
(\psi\psi^\dagger)_{jj}(\psi_{k,-\chi})_i\tilde{c}^\dagger_{is'}d_{ls'\chi}
|G_{s\bar{s}}\rangle\bigr|_{j\in\circ}\nonumber\\
&\phantom{=}
+\frac{3}{4}J\sum_{j\in\circ}(\psi\psi^\dagger)_{ii}
(\psi_{l\chi})_j^\ast d^\dagger_{ks',-\chi}\tilde{c}_{js'}
|G_{s\bar{s}}\rangle\bigr|_{i\in\bullet}\\
&=\frac{3}{2}J\phi|E_{ss'\chi}^{kl}\rangle\label{eq:gap-exchange},
\end{align}
% \begin{gather}
% \tilde{c}_{i\downarrow}\tilde{c}_{i\uparrow}|G_{s\bar{s}}\rangle
% =\tilde{c}^\dagger_{i\uparrow}\tilde{c}^\dagger_{i\downarrow}|G_{s\bar{s}}\rangle\\
% \tilde{c}^\dagger_{j\bar{s}}\tilde{c}_{is}|G_{s\bar{s}}\rangle
% \end{gather}
where we have exploited the anticommutation relations,
\begin{gather}
 \{\tilde{c}_{is},\tilde{c}^\dagger_{js'}\}=(\psi\psi^\dagger)_{ij}\delta_{ss'},\qquad
\{\tilde{c}_{is},d^\dagger_{ks'\chi}\}=(\psi_{k\chi})_i\delta_{ss'},\\
\{\tilde{c}_{is},\tilde{c}_{js'}\}
=\{\tilde{c}^\dagger_{is},\tilde{c}^\dagger_{js'}\}=0.
\end{gather}
Combining Eqs.~(\ref{eq:gap-on-site}) and (\ref{eq:gap-exchange}),
we arrive at the expression for the charge gap~(\ref{eq:charge-gap})
that is linearly dependent on $\phi$.

\bibliography{references}

\end{document}